\documentclass[prd,twocolumn,amsfonts,showpacs]{revtex4}

\newcommand{\beq}{\begin{equation}}
\newcommand{\eeq}{\end{equation}}
\newcommand{\beqs}{\begin{eqnarray}}
\newcommand{\eeqs}{\end{eqnarray}}
\newcommand{\lsim}{\mathrel{\raisebox{-
.6ex}{$\stackrel{\textstyle<}{\sim}$}}}

\newcommand{\pslash}{p\hspace{-0.067in}\slash}

\begin{document}

\title{Maximum Wavelength of Confined Quarks and Gluons and Properties 
of Quantum Chromodynamics}

\author{Stanley J. Brodsky$^{a,b,c}$}

\author{Robert Shrock$^b$}

\affiliation{(a)
Stanford Linear Accelerator Center, Stanford University, Stanford, CA  94309}

\affiliation{(b)
C.N. Yang Institute for Theoretical Physics, Stony Brook University,
Stony Brook, NY 11794}

\affiliation{(c)
Department of Physics, University of Durham, Durham DH1 3LE, UK}

\begin{abstract}

Because quarks and gluons are confined within hadrons, they have a maximum
wavelength of order the confinement scale. Propagators, normally calculated for
free quarks and gluons using Dyson-Schwinger equations, are modified by
bound-state effects in close analogy to the calculation of the Lamb shift in
atomic physics.  Because of confinement, the effective quantum chromodynamic
coupling stays finite in the infrared.  The quark condensate which arises from
spontaneous chiral symmetry breaking in the bound state Dyson-Schwinger
equation is the expectation value of the operator $\bar q q$ evaluated in the
background of the fields of the other hadronic constituents, in contrast to a
true vacuum expectation value. Thus quark and gluon condensates reside within
hadrons.  The effects of instantons are also modified.  We discuss the
implications of the maximum quark and gluon wavelength for phenomena such as
deep inelastic scattering and annihilation, the decay of heavy quarkonia, jets,
and dimensional counting rules for exclusive reactions.  We also discuss
implications for the zero-temperature phase structure of a vectorial SU($N$)
gauge theory with a variable number $N_f$ of massless fermions.

\end{abstract} 


\maketitle

\section{Introduction}

Bethe's remarkable calculation of the Lamb shift in hydrogen in 1947
\cite{bethe} laid the foundation for the renormalization procedure in quantum
field theory and the subsequent development of quantum electrodynamics (QED).
The Lamb shift is the change in the bound-state electron energy, in particular,
the $2S_{1/2}$ and $2P_{1/2}$ levels of hydrogen, as a result of the effect of
fluctuations in the quantized electromagnetic field on the electron.  An
essential aspect of the Lamb shift calculation in QED is the fact that while
the wavefunction renormalization constant of a free electron $Z_2$ is infrared
(IR) divergent, it becomes infrared-finite when the electron is bound in an
atom.  In the case of a free electron, the IR divergences are cancelled when
one properly considers electron propagation together with the emission of soft
real photons. In the case of the electron in an atomic bound state, the $k$
integration over photon momenta is cut off in the infrared by the fact that the
relevant photon wavelengths have a maximum value set by the size of the atom,
i.e., $k \ge k_{min,atom}$,
\beq
k_{min,atom} \simeq \frac{1}{na_B} \simeq \frac{\alpha_{em} m_e}{n} \ , 
\label{kminatom}
\eeq
where $a_B$ is the Bohr radius and $n$ is the radial quantum number
characterizing the bound state of the electron in the Coulomb field of the
proton.  After mass renormalization photon momenta larger than $k \simeq m$ do
not appreciably affect the electron motion. Combining these cutoffs then yields
the Bethe logarithm, $\ln(1/\alpha_{em})$ in the energy shift \cite{bethe}.

The complete calculation of energy levels of Coulombic bound states in QED,
such as hydrogen, positronium, or muonium ($\mu^+e^-$), begins with the
Bethe-Salpeter equation for the two bound-state particles.  In the case where
one of these particles can be taken as very heavy, such as in hydrogen or
muonium, the bound-state electron propagator is replaced by the resolvent
\beq
\frac{1}{\Pi \cdot \gamma - m_e+ i \epsilon} \ , 
\label{resolve}
\eeq
where $\Pi^\mu = p^\mu - e A^\mu$ and $A^\mu$ is the background electromagnetic
field of the heavy particle.  An analysis of the bound-state electron
self-energy in terms of a gauge-invariant expansion in the electromagnetic
field strength of the background field is given in ref. \cite{ybe}.

Quantum chromodynamics (QCD) has provided a remarkably successful theory of
hadrons and strong interactions. At short distance; i.e., large Euclidean
momentum scales $\mu$, the squared QCD gauge coupling, $\alpha_s(\mu) =
g_s^2(\mu)/(4\pi)$ becomes small, as a consequence of asymptotic freedom.  As
the momentum scale decreases through $\Lambda_{em} \simeq 200$ MeV, the theory
exhibits spontaneous chiral symmetry breaking, and quarks and gluons are
confined within color-singlet physical states, the hadrons.  Thus, because of
confinement, quarks and gluons have maximum wavelengths
\beq
\lambda_{max} \simeq \Lambda_{QCD}^{-1} \simeq 1 \ {\rm fm} \ . 
\label{lambdamax}
\eeq
Equivalently, the quantum mechanical wavefunctions for quarks and
gluons within hadrons have minimum bound-state momenta, 
\beq
k_{min} = |{\bf k}|_{min} \simeq \Lambda_{QCD} \ . 
\label{kminhadron}
\eeq 
For example, in the hard-wall model of AdS/QCD, color confinement of quarks in
the AdS fifth dimension gives the frame-independent condition at equal
light-front time~\cite{Brodsky:2008gc}:
\beq
\zeta =\sqrt{b^2_\perp  x(1-x)} < 
\lambda_{max} 
\label{lambdamaxAdS}
\eeq
where $\vec b_\perp$ is the quark-antiquark impact separation and $x$ is the
quark light-cone momentum fraction $x = k^+/P^+ =(k^0 + k^3)/(P^0 + P^3).$ Thus
in principle all perturbative and nonperturbative QCD analyses should be
performed with the infrared regularization imposed by color confinement.

Here we point out and discuss several important consequences of color
confinement and the resultant maximum wavelength of quarks and gluons that do
not seem to have received attention in the literature.  These include
implications for an infrared fixed point of the QCD $\beta$ function and new
insights into spontaneous chiral symmetry breaking which are apparent when one
uses bound-state rather than free Dyson-Schwinger equations (DSE's) in close
analogy to QED bound state computations. An important consequence is that quark
and gluon condensates are confined within hadrons, rather than existing
throughout space-time. We also discuss hadron mass calculations using
Bethe-Salpeter equations (BSE's), analyze the modifications of instanton
physics, and comment on insights that one gains concerning
short-distance-dominated processes and dimensional counting for hard exclusive
processes.

Parenthetically, we note that if QCD contains $N_f \ge 2$ exactly massless
quarks and if one turned off electroweak interactions, then the theory would
have a resultant set of $N_f^2-1$ exactly massless Nambu-Goldstone bosons
(e.g., the pions, for the case $N_f=2$). Because the size of a nucleon, $r_N$,
is determined by the emission and reabsorption of virtual pions and the
resultant pion cloud, and hence is $r_N \sim m_\pi^{-1}$, this nucleon size
would be much larger than $\Lambda_{QCD}^{-1}$.  Since in the real world,
$m_\pi$ is not $<< \Lambda_{QCD}$, we do not pursue the analysis of this {\it
gedanken} world here.

\section{Implications of $\lambda_{max}$ for the Infrared Behavior of QCD} 

The fact that quarks and gluons have maximum wavelengths $\lambda_{max}$ has
important consequences for the infrared behavior of QCD.  For $\mu^2 >>
\Lambda_{QCD}^2$, QCD is weakly coupled, and the evolution of $\alpha_s$ is
described by the $\beta$ function
\beq
\beta(t) = \frac{d\alpha_s}{dt} = - \frac{\alpha_s^2}{2\pi}\bigg ( b_1 + 
\frac{b_2 \alpha_s}{4\pi} + O(\alpha_s^3) \bigg ) \ , 
\label{beta}
\eeq
where $t=\ln\mu$ and the one- and two-loop coefficients $b_\ell$, $\ell=1,2$
are scheme-independent, while higher-order coefficients are scheme-dependent.

In the standard perturbative calculations of these coefficients, one performs
integrations over Euclidean loop momenta ranging from $k=0$ to $k=\infty$.
Although this is a correct procedure for describing the evolution of
$\alpha_s(\mu)$ in the ultraviolet region $\mu >> \Lambda_{QCD}$ where the
coupling is weak and effects of confinement are unimportant, it does not
incorporate the property of confinement at low scales $\mu \lsim
\Lambda_{QCD}.$ Confinement implies that both the gluons and quarks have
restricted values of momenta $k \ge k_{min}.$ Since loop corrections to the
gluon propagator vanish as $q^2/k_{min}^2$ when this ratio is small, it follows
that (to the extent that one can continue to use the quark and gluon fields and
the associated coupling $\alpha_s$ to describe the physics in this region of
momenta), as $q^2$ decreases in magnitude below $\Lambda_{QCD}^2$, the QCD
$\beta$ function measuring the evolution of $\alpha$ at the scale $\mu^2 \simeq
q^2$ must also vanish in the infrared.  This implies a physical cutoff on the
growth of $\alpha_s(\mu)$ at small $\mu^2$; i.e., infrared fixed-point behavior
of the QCD coupling.  In effect, QCD exhibits a well-defined limiting behavior
in the infrared, and the infrared growth of $\alpha_s$ is suppressed, not
because the perturbative $\beta$ function exhibits a zero away from the origin,
but because confinement provides an infrared cutoff.  In fact, as summarized in
reference~\cite{Deur:2008rf}, effective QCD couplings measured in experiments,
such as the effective charge $\alpha_{s,g_1}(Q^2)$ defined from the Bjorken sum
rule, display a lack of $Q^2$-dependence in the low $Q^2$ domain.

\section{Implications for QCD Phenomenology}

The Bjorken scaling of the deep inelastic lepton-nucleon scattering (DIS) $\ell
N \to \ell' X$ cross section is explained in perturbative QCD by arguing that
the emission of gluons from the scattered quark is governed by a coupling
$\alpha_s$ which is small because of asymptotic freedom.  Thus in the
leading-twist DIS regime, the DIS structure functions $F_i(x,Q^2)$ are mainly 
functions of $x$, with only small, logarithmic dependence on $Q^2$.  A similar
argument underlies the application of perturbative QCD and the parton model to
deep inelastic annihilation, $e^+e^- \to$ hadrons with center-of-mass energy
squared $s >> \Lambda_{QCD}^2$, away from thresholds, leading to the formula
$\sigma(e^+e^- \to \ {\rm hadrons})/\sigma(e^+e^- \to \mu^+\mu^-) = R$ with
$R(s) = N_c \sum_j q_j^2$, where the sum is over quarks with $4m_q^2 < s$.
Asymptotic freedom in QCD has also been used to explain the narrow widths of
${}^3S_1$, $J^{PC}=1^{--}$ $Q \bar Q$ states of heavy (i.e., $m_Q >>
\Lambda_{QCD}$) quarks with masses below threshold for emission of the
associated heavy-flavor mesons.  The explanation is, in essence, that hadronic
decays proceed by emission of three gluons, and because the corresponding
couplings $g_s(\mu)$ are small for $\mu \sim m_Q/3$ due to asymptotic freedom,
the resultant decays are suppressed \cite{ap}.  A fourth property of QCD which
makes use of its property of asymptotic freedom is the phenomenon of jets
\cite{gsjets}.

In each of these examples, the theoretical analyses provide successful
descriptions of the physical processes in terms of the asymptotic freedom of
QCD. The analyses involve a factorization between the short-distance,
perturbatively calculable, part of the process and the long-distance part
involving hadronization.  However, it is important to ask whether such
calculations are stable against multiple gluon emission. For example, if the
outgoing struck quark in DIS or the $q \bar q$ pair in DIA were to radiate a
sufficiently large number of gluons, then, since on average each of these would
carry only a small momentum, the associated coupling $\alpha_s$ would not be
small, and this could significantly change the prediction for the cross
section.  A related concern regards the narrow width of orthoquarkonium: if the
heavy $Q$ and $\bar Q$ annihilate not into three gluons, but into a
considerably larger number, $\ell$, of gluons, then each would carry a much
smaller momentum, $k \sim 2m_Q/\ell$, and thus the associated running coupling
$g_s(k)$ would not be small.  Similarly, in a hard scattering process that
leads to the production of a $q \bar q$ pair with invariant mass $\hat s = (p_q
+ p_{\bar q})^2 >> \Lambda_{QCD}^2$ and $\hat s >> 4m_q^2$, the $q$ and $\bar
q$ might radiate a large number of gluons, each carrying a small relative
momentum; again the resulting running coupling is consequently not small, and
this could dilute the jet-like structure of the event.  A similar concern
applies to jets involving gluons.

Here we provide a simple physical explanation of why gluon emission is
stabilized: because of confinement, the gluons have minimum momenta of order
$\Lambda_{QCD}$, so the apparently dangerous scenario involving emission of a
large number of gluons with momenta of order $\Lambda_{QCD}$ or smaller is
kinematically impossible.

An infrared cutoff on the growth of the QCD coupling also helps to explain the
remarkable success of dimensional counting rules in describing data on the
differential cross sections of exclusive reactions at high-energy and
fixed-angle in QCD \cite{dimcount}.  Recall that for a reaction $a + b \to c_1
+ ... + c_k $ with $s >> \Lambda_{QCD}^2$, $t=Q^2 >> \Lambda_{QCD}^2$, and
fixed $s/t$ (i.e., fixed CM scattering angle), dimensional counting gives
\cite{dimcount}
\beq
\frac{d\sigma}{dt} \propto s^{-(n-2)} \ , 
\label{dsigmadt}
\eeq
where the twist $n$ denotes the total number of elementary valence fields
entering the hard scattering amplitude.  This implies, for example, that, under
these conditions, $d\sigma(pp \to pp)/dt \propto s^{-10}$, $d\sigma(\pi p \to
\pi p) \propto s^{-8}$, etc.  One might worry that since all of the external
particles in these reactions are on-shell, eq. (\ref{dsigmadt}) might receive
large long-distance corrections. An appealing explanation for the absence of
such corrections is that they are suppressed by the cutoff in the growth of the
running QCD coupling $\alpha_s$ due to the $k_{min}$ of the gluons.  This is
also consistent with fits to data \cite{bggr,field02,btrev}.

\section{Implications for Spontaneous Chiral Symmetry Breaking}

A limit on the maximum of gluon and quark wavelengths also has implications for
spontaneous chiral symmetry breaking (S$\chi$SB) in QCD.  It is clear that
because of confinement, analyses based on free quark and gluon propagators,
such as the Dyson-Schwinger equation, need to be replaced in principle by
analyses which incorporate bound-state dynamics, such as the QCD Bethe-Salpeter
equation.  Recall that since the $u$ and $d$ current-quark masses are $<<
\Lambda_{QCD}$, the QCD Lagrangian theory has a global ${\rm SU}(N_f)_L \times
{\rm SU}(N_f)_R$ chiral symmetry, broken spontaneously to the diagonal, vector
isospin subgroup ${\rm SU}(N_f)_{diag}$, where $N_f=2$, by the $\langle \bar q
q \rangle$ condensates with $q=u,d.$ (The analogous statement applies to the
corresponding symmetry with $N_f=3$, with larger explicit breaking via $m_s$.)
Some studies of spontaneous chiral symmetry breaking in QCD are listed in
ref. \cite{ds}.

The inverse quark propagator has the form $S_f(p)^{-1} = A(p^2) \pslash -
B(p^2)$. In the one-gluon exchange approximation, the DSE for $S_f^{-1}$ is
\beq
S_f(p)^{-1} - \pslash = -i C_F g^2 \int \frac{d^4 k}{(2\pi)^4}
\, D_{\mu\nu}(p-k) \,  \gamma^\mu \, S_f(k) \, \gamma^\nu
\label{sdeq}
\eeq
where $C_F$ is the quadratic Casimir invariant and $D_{\mu\nu}(k)$ is the gluon
propagator.  Chiral symmetry breaking is a gauge-invariant phenomenon, so one
may use any gauge in solving this equation.  It is convenient to use the Landau
gauge since then there is no fermion wavefunction renormalization; i.e.,
$A(p^2)=1$.  Equation (\ref{sdeq}) has a nonzero solution for the dynamically
generated fermion mass $\Sigma$ (which can be taken to be $\Sigma(p^2) =
B(p^2)$ for Euclidean $p^2 << \Lambda^2$) if $\alpha_s \ge \alpha_{cr}$,
where $3 \alpha_{cr} C_F/\pi = 1$.  Since $\Sigma$ is formally a source in
the path integral for the operator $\bar qq$, one associates $\Sigma \ne 0$
with a nonzero quark condensate.  Clearly, this only provides a rough estimate
of $\alpha_{cr}$, in view of the strong-coupling nature of the physics and
the consequent large higher-order perturbative, and also nonperturbative,
contributions.

If one now takes quark and gluon confinement into account, then just as for the
Lamb shift, the integral over loop momenta in eq. (\ref{sdeq}) can extend in
the infrared only to $k_{min} \sim \Lambda_{QCD}$.  Although the DSE analysis
for the free quark propagator may incorporate some of the physics relevant to
spontaneous chiral symmetry breaking, it does not incorporate the property of
confinement. This is an important omission since a plausible physical
explanation for spontaneous chiral symmetry breaking in QCD involves
confinement in a crucial manner; this breaking results from the reversal in
helicity (chirality) of a massless quark as it heads outward from the center of
a hadron and is reflected back at the outer boundary of the hadron
\cite{casher}.

The Dyson-Schwinger equation has been used in conjunction with the
Bethe-Salpeter equation for approximate calculations of hadron masses and other
quantities in QCD \cite{bs}.  Our observation implies that here again, the
integration over virtual loop momenta in the BSE can only extend down to
$k_{min} \sim \Lambda_{QCD}$, not to $k=0$.  This obviates the need for
artificial cutoffs on the growth of the QCD coupling occurring in the integrand
that have been employed in past studies.  Analyses using the DSE and BSE have
been used to calculate hadron masses in the confining phase of an abstract
asymptotically free, vectorial SU($N$) gauge theories with a variable number,
$N_f$, of light fermions \cite{bsnf}.  It would be worthwhile to incorporate
the effect of $k_{min}$ in these studies, as well as in analyses for actual
QCD.

Thus let us consider the propagator of a light quark bound in a light-heavy $q
\bar Q$ meson, such as $B^+=(u\bar b)$ or $B^0_d = (d \bar b)$.  At
sufficiently strong coupling $\alpha_s$, the DSE (in this case, effectively a
bound-state Dyson-Schwinger equation) yields a nonzero, dynamically generated
mass, $\Sigma$ for the light quark.  One can associate this with a bilinear
quark condensate $\langle \bar q q \rangle$, as noted above.  However, this
condensate is the expectation value of the operator $\bar q q$ in the
background (approximately Coulombic) field of the heavy $\bar b$ antiquark, in
contrast to a true vacuum expectation value \cite{cond}.  This is in accord
with our argument in \cite{cond} that the quark condensate $\langle \bar q q
\rangle$ and gluon condensate $\langle {\rm Tr}(G_{\mu\nu}G^{\mu\nu})\rangle$
have spatial support only in the interior of hadrons, since that is where the
quarks and gluons which give rise to it are confined.  In \cite{cond} we noted
that this conclusion is the analogue for quantum field theory of the
experimental fact that the spontaneous magnetization below the Curie
temperature in a piece of iron has spatial support only within the iron rather
than extending to spatial infinity. Our observation in Ref. \cite{cond} that
QCD condensates have spatial support restricted to the interiors of hadrons has
the important consequence that these condensates contribute to the mean baryon
mass density in the universe, but not to the cosmological constant or dark
energy \cite{cc}.  As in the case of QED, the bound-state problem of a light
quark in a background field can be formulated as a resolvent problem or in
terms of an effective theory \cite{Brambilla:2004jw}.  Of course, in the
hypothetical situation in which the current-quark masses $m_u=m_d=0$ and one
turned off electroweak interactions, so that $m_\pi \to 0$, the size of
hadrons, as determined by their meson clouds, would become infinitely large,
and the condensates would have infinite spatial extent.

In general, the breaking of a continuous global symmetry gives rise to
Nambu-Goldstone modes.  It was noted in Ref. \cite{cond} that in a sample of a
ferromagnetic material below its Curie temperature, these Nambu-Goldstone spin
wave modes (magnons) are experimentally measured to reside within the sample.
Moreover, via the correspondence between the partition function defining a
statistical mechanical system and the path integral defining a quantum field
theory, there is an analogy between the spin waves in a Heisenberg ferromagnet
and the almost Nambu-Goldstone modes in QCD - the pions.  As required for the
self-consistency of our analysis of condensates, the wavefunctions for the
pions have spatial support where the chiral condensates exist, namely within
the pions themselves and, via virtual emission and reabsorption, within other
hadrons, in particular nucleons.

  We relate these statements to some current algebra results.  Consider the
vacuum-to-vacuum correlator of the axial-vector current, $\langle
0|J^\mu_5(x_a) J^\mu_5(x_b)|0 \rangle$.  The Fourier transform of the cut of
this propagator can, in principle, be measured in $e^+ e^- \to Z^*_0$
axial-vector current events. The pion appears as a pole in this propagator at
$q^2 = m^2_\pi$, corresponding to $e^+ e^- \to Z^*_0\to \pi^0$.  Here the
axial-vector current $J^\mu_5 = \bar q \gamma^\mu \gamma^5 q$ creates a quark
pair at $x_a$ which propagates to $x_b$, and as the $q$ and $ \bar q$
propagate, they interact and bind to create the pion.  For example, at fixed
light-front time $x_a^+=x_b^+$, the pion pole contribution appears when $-(x_a
- x_b)^2=(x^\perp_a-x^\perp_b)^2 \simeq R^2_\pi$,where $R_\pi$ is the
transverse pion size, of order $1/m_\pi.$ The axial current that couples to the
pion thus involves $\bar q q$ creation within a domain of the size of the pion.
Furthermore, any $q\bar q$ condensate that appears in the quark or antiquark
propagator in this process is created in a finite domain of the pion size,
since that is where the quark and antiquark propagate, subject to the confining
color interaction.  Again, we see that one needs to use a modified form of the
DSE for the quark propagator which takes into account the field of the
antiquark.  And again, the loop momenta have a maximum wavelength corresponding
to the finite size of the domain where color can exist.

Since the property that quarks and gluons have a maximum wavelength is a
general consequence of confinement, it necessarily appears in specific
phenomenological models of confined hadrons, such as bag models \cite{mitbag}
(see also \cite{continuum}) and recent approaches using AdS/CFT methods
\cite{btrev}.  It is also evident in lattice gauge simulations of QCD
\cite{lgtgluons}.  Sometimes the results of these simulations are phrased as
the dynamical generation of an effective ``gluon mass''; here we prefer to
describe the physics in terms of a maximum gluon wavelength, since this
emphasizes the gauge invariance of the phenomenon.

\section{Implications for Instanton Effects in QCD}

The maximum wavelength of gluons also has implications for the role of
instantons in QCD.  In the semiclassical picture, after a Euclidean rotation,
one identifies the gluon field configurations which give the dominant
contribution to the path integral as those which minimize the Euclidean action.
This requires that the field strength tensor $F_{\mu\nu} \equiv \sum_a T_a
F^a_{\mu\nu}$ vanish as $|x| \to \infty$, which implies that $A_\mu =
-(i/g_s)(\partial_\mu U)U^{-1}$, where $A_\mu \equiv \sum_a T_a A^a_\mu$ and
$U(x) \in {\rm SU}(N_c)$.  Since the outer boundary of the compactified
${\mathbb R}^4$ in this $|x| \to \infty$ limit is $S^3$, the gauge fields thus
fall into topologically distinct classes, as described by the class of
continuous mappings from $S^3$ to SU($N_c$), i.e., the homotopy group
$\pi_3({\rm SU}(N_c))={\mathbb Z}$.  Instantons appear to play an important
role in QCD, explaining, for example, the breaking of the global U(1)$_A$
symmetry and the resultant fact that the $\eta'$ meson is not light
\cite{hooft,cdg}; however, one should recognize that, because of confinement,
the gauge fields actually have no support beyond length scales of order
$1/\Lambda_{QCD} \sim 1$ fm.  The semiclassical analysis with its $|x| \to
\infty$ limit used to derive the result $A_\mu = -(i/g_s)(\partial_\mu
U)U^{-1}$ is thus subject to significant corrections due to confinement.  This
is evident in the BPST instanton solution for $N_c=2$, namely \cite{bpst}
\beq
ig_s A_\mu = \bigg ( \frac{x^2}{x^2+\rho^2} \bigg ) (\partial_\mu U)U^{-1} \ ,
\label{a_bpst}
\eeq
where $U(x) = (x^0 + i \tau \cdot {\bf x})/|x|$.  This form only becomes a pure
gauge in the limit $|x| \to \infty$.  It has long been recognized that in
calculating effects of instantons in QCD, which involve integrations over
instanton scale size $\rho$, uncertainties arise due to the fact that there are
big contributions from instantons with large scale sizes, where the
semiclassical approximation is not accurate \cite{cdg}.  Our point is
different, although related; namely that the accuracy of the semiclassical
instanton analysis is also restricted by the property that $\lambda_{max} \sim
1$ fm for the gluon field and hence one cannot really take the $|x| \to \infty$
limit in the manner discussed above.

\section{Implications for General Non-Abelian Gauge Theories}

Our observations are also relevant to the problem of determining the phase
structure (at zero temperature and chemical potential) of a vectorial SU($N$)
gauge theory with a gauge coupling $g$ and a given content of massless
fermions, such as $N_f$ fermions transforming according to the fundamental
representation of SU($N$).  We assume $N_f < (11/2)N$, so that the theory is
asymptotically free.  Since fermions screen the gauge field, one expects that
for sufficiently large $N_f$, the gauge interaction would be too weak to
confine or produce spontaneous chiral breaking (e.g., \cite{bz}).  An estimate
of the critical value, $N_{f,cr}$, beyond which there would be a phase
transition from a phase with confinement and S$\chi$SB to one without such
symmetry breaking (and presumably without confinement) has been obtained
combining the perturbative $\beta$ function and the DSE.  We denote this as the
$\beta$DS method.  For sufficiently large $N_f$, the perturbative
$\beta$ function exhibits a zero away from the origin, at $\alpha_{\rm IR}$,
where $\alpha=g^2/(4\pi)$. The value of $\alpha_{\rm IR}$ is a decreasing
function of $N_f$.  The value of $N_{f,cr}$ is determined by the condition that
$\alpha_{IR}$ decreases below the minimal value $\alpha_{cr}$ for which the
approximate solution to the DS equation yields a nonzero solution for the
dynamically generated fermion mass.  The $\beta$DS analysis, to two-loop
accuracy, yields the estimate \cite{chipt}
\beq
N_{f,cr} = (N_{f,cr})_{2\ell \beta {\rm DS}} = \frac{2N(50N^2-33)}{5(5N^2-3)} \ ,
\label{nfcr}
\eeq
where $2\ell$ refers to the two-loop accuracy to which the beta function is
calculated. This gives $N_{f,cr} \simeq 8$ for $N=2$ and $N_{f,cr} \simeq 12$
for $N=3$.  The value of $N_{f,cr}$ is important because if the approximate IR
fixed point $\alpha_{IR}$ is larger than, but near to, $\alpha_{cr}$, the
resultant gauge coupling runs slowly over an extended interval of energies.
This ``walking" behavior is useful for models of dynamical electroweak symmetry
breaking \cite{chipt,wtc}.  (In such models there are motivations for choosing
$N=2$, including a mechanism to explain light neutrino masses \cite{nt}.)  The
effect of the three-loop terms in the $\beta$ function and the next higher-loop
terms in the DSE have been studied in Ref. \cite{alm}.

Neither the $\beta$ function nor the DSE used in this $\beta$DS method
includes the effect of instantons.  Studies in QCD have shown that instantons
enhance spontaneous chiral symmetry breaking \cite{ccdg}.  Analyses of
instanton effects on fermion propagation were carried out for general $N_f$,
and these were shown to contribute substantially to S$\chi$SB
\cite{instantons}.  One would thus expect that if one augmented the $\beta$DS
approach to include effects of instantons, the resultant improved estimate of
$N_{f,cr}$ would be greater than the value obtained from the $\beta$DS method
without instantons.

In principle, lattice gauge theory can provide a fully nonperturbative approach
for calculating $N_{f,cr}$ \cite{iwasaki}-\cite{afn}.  One lattice group has
obtained values of $N_{f,cr}$ considerably smaller than the respective
$\beta$DS values (for $N=2,3$) \cite{iwasaki}; however, the most recent study
of the $N=3$ case finds evidence that the infrared behavior of the theory is
conformal for $N_f \ge 12$ but exhibits confinement and chiral symmetry
breaking for $N_f \le 8$, consistent with the $\beta$DS analysis
\cite{afn}. Since the $\beta$DS method does not include instanton effects,
there is thus a question why it appears to produce a rather accurate value of
$N_{f,cr}$.

Our observation provides a plausible answer to this question.  Approaching the
chiral boundary from within the phase with confinement and spontaneous chiral
symmetry breaking, we note that the confinement-induced $k_{min}$ of the gluons
reduces their contribution to the increase of the gauge coupling in the
infrared and also to the virtual gluon exchange effects on the fermion
propagator.  This reduction of gluonic effects acts in the opposite direction
relative to the enhancement of chiral symmetry breaking due to instantons, and
thus has the potential to explain why the $\beta$DS estimate for $N_{f,cr}$,
which does not incorporate either confinement or instanton effects, could
nevertheless yield a reasonably accurate value for $N_{f,cr}$.  Moreover, as
noted above, the role of instantons in S$\chi$SB is affected by the confinement
of gluons and the resultant corrections to the semiclassical approach to QCD.

\section{Conclusions}

Because quarks and gluons are confined within hadrons, there is maximum limit
on their wavelengths. Propagators, normally calculated for free quark and
gluons using Dyson-Schwinger equations, are modified in the infrared by bound
state effects in close analogy to the calculation of the Lamb shift in atomic
physics.  Thus because of confinement, the effective QCD coupling stays finite
and flat at low momenta.  The quark condensate which arises from spontaneous
chiral symmetry breaking in the bound-state Dyson-Schwinger equation is the
expectation value of the operator $\bar q q$ evaluated in the background of the
fields of the other hadronic constituents, in contrast to a true vacuum
expectation value. Thus quark and gluon condensates have support only within
hadrons.

We have shown that the limit on the maximum wavelength of gluons and quarks
from confinement leads to new insights into a number of phenomena in QCD,
including deep inelastic scattering and annihilation, the narrow widths of
heavy orthoquarkonium states, jets, instantons, dimensional counting rules for
hard exclusive processes, and other phenomena related to the infrared behavior
of the theory.  We have also given a plausible explanation of how estimates of
the value $N_{f,cr}$ in a general asymptotically free, vectorial SU($N$) theory
based on a method using the perturbative $\beta$ function and the
Dyson-Schwinger equation could be reasonably accurate even though this method
does not incorporate the effects of confinement or instantons.  Our
observations suggest a program of future research devoted to incorporating the
effect of the maximum wavelength of quarks and gluons in analytic studies of
QCD properties.

\section{Acknowledgments}

We are grateful to Guy de Teramond for helpful discussions and comments. This
research was partially supported by grants DE-AC02-76SF00515 (SJB) and
NSF-PHY-06-53342 (RS).  Preprint SLAC-PUB-13246, IPPP/08/37, DCPT/08/74,
YITP-SB-08-11.

\end{document}